\documentclass[10pt, conference]{IEEEtran}

\IEEEoverridecommandlockouts
\usepackage{cite}
\usepackage{amsmath,amssymb,amsfonts,amsthm}
\usepackage{graphicx}
\usepackage{textcomp}
\usepackage{xcolor}
\usepackage{svg}
\usepackage{algorithm}
\usepackage{algorithmic}
\usepackage{balance}
\usepackage{multirow}
\usepackage{fancyhdr}

\def\BibTeX{{\rm B\kern-.05em{\sc i\kern-.025em b}\kern-.08em
    T\kern-.1667em\lower.7ex\hbox{E}\kern-.125emX}}

\newtheorem{lemma}{Lemma}

\begin{document}

\title{Optimization in Mobile Augmented Reality Systems for the Metaverse over Wireless Communications
}

\author{
 \IEEEauthorblockN{Tianming Lan, Jun Zhao}

 \IEEEauthorblockA{School of Computer Science and Engineering \\ Nanyang Technological University, Singapore}
 \IEEEauthorblockA{tianming001@e.ntu.edu.sg, junzhao@ntu.edu.sg}
}
\maketitle
 \thispagestyle{fancy}
\pagestyle{fancy}
\lhead{This paper appears in IEEE Global Communications Conference (GLOBECOM) 2023.}
\cfoot{\thepage}
\renewcommand{\headrulewidth}{0.4pt}
\renewcommand{\footrulewidth}{0pt}

\begin{abstract}
As the essential technical support for Metaverse, Mobile Augmented Reality (MAR) has attracted the attention of many researchers. MAR applications rely on real-time processing of visual and audio data, and thus those heavy workloads can quickly drain the battery of a mobile device. To address such problem, edge-based solutions have appeared for handling some tasks that require more computing power. However, such strategies introduce a new trade-off: reducing the network latency and overall energy consumption requires limiting the size of the data sent to the edge server, which, in turn, results in lower accuracy.
In this paper, we design an edge-based MAR system and propose a mathematical model to describe it and analyze the trade-off between latency, accuracy, server resources allocation and energy consumption. Furthermore, an algorithm named LEAO is proposed to solve this problem.
We evaluate the performance of the LEAO and other related algorithms across various simulation scenarios. The results demonstrate the superiority of the LEAO algorithm.
Finally, our work provides insight into optimization problem in edge-based MAR system for Metaverse.
\end{abstract}


\section{Introduction}
Metaverse has become more  and more popular nowadays. And Augmented Reality (AR) is one of Metaverse's important technical supports \cite{laviola20173d}, which can integrate the digital and physical elements \cite{tuan2018demonstration}. Mobile Augmented Reality (MAR) can be more convenient than AR since people can use mobile devices to get access to the Metaverse everywhere. 

To integrate MAR applications into the Metaverse, accurate and real-time recognition is crucial, because recognition failure and high latency will deteriorate the user experience \cite{lee2020ubipoint}.
However, on mobile devices, MAR applications, such as Target Recognition \cite{redmon2016you}, are too energy-intensive to use. And limited computing resources will result in the long calculation latency.
Hence, some research tried to offload image recognition tasks to the server that has strong computing power \cite{jain2016low, chen2015glimpse, shea2017towards}.\par
However, the use of servers, such as edge servers \cite{shi2016edge, satyanarayanan2017emergence}, can result in long transmission delays and unstable networks, which may lead to poor user experience. Therefore, trade-offs become particularly important in such scenarios. Many researchers have begun to study the trade-off in this scenario which they call edge-based MAR systems \cite{liu2019edge, wang2020user}.\par
\textbf{Motivation}. On the one hand, when mobile devices send image recognition tasks to the edge server, high-resolution images could lead to high training accuracy, but low-resolution images could save transmission energy and latency. Therefore, there will be a trade-off between energy, latency and accuracy. On the other hand, with the emergence of the Metaverse, the rapid growth of user and data volume will bring enormous pressure to edge servers. How to allocate the resources of multiple edge servers to all users in a reasonable manner while achieving the above trade-off is also an urgent and challenging issue to be addressed.\par

\textbf{Challenges}. This paper takes into account all the factors to the best of its ability, including latency, allocation of multi-server resources, accuracy, and user energy consumption. We also take the server energy consumption as one of the optimization goals because few research considered it. The aforementioned trade-off is formulated as a Mixed-Integer Non-convex Problem (MINP), which is generally known to be NP-hard. Such problems can not be easily solved with conventional optimization methods. This paper employs various mathematical techniques and proposes the \underline{L}atency, \underline{E}nergy consumption, resources allocation and \underline{A}ccuracy comprehensive \underline{O}ptimization algorithm (LEAO) algorithm to address this issue.

After the comparison with other methods, we conclude that LEAO has a good performance in this scenario.\par

Our contributions are summarized as follows: \par
\begin{enumerate}
\item We find the trade-off problem between latency, accuracy, resources allocation and energy consumption in MAR scenario.We also take server energy consumption into account firstly. Finally, we propose an analytical model to formulate this problem.
\item We propose the LEAO algorithm to solve the trade-off with various mathematical techniques.
\item We design a MAR system based on our analytical model and we conduct experiments to verify the performance of LEAO in this system.
\end{enumerate}
The organization of this paper is as follows: Section \ref{Related work} will introduce the related work; Section \ref{Problem Formulation} will describe the analytical model and how we formulate the problem; Section \ref{Algorithm} will introduce optimization algorithm detail; Section \ref{simulation} will introduce our experimental methods and results; Section \ref{Conclusion} will summarize this paper.

\section{Related work}\label{Related work}
The attempts at solving above-mentioned trade-offs start with cloud-based MAR systems \cite{chatzopoulos2017mobile}. Chen $et \ al.$ proposed a real-time object recognition system \cite{chen2015glimpse}. This system can improve accuracy and select appropriate image pixels automatically. Jain $et \ al.$ proposed a method to reduce network latency \cite{jain2016low}. Liu $et \ al.$ proposed FACT algorithm to find the optimal point between accuracy and network latency \cite{liu2018edge}. But this work ignores the energy consumption in MAR system, which is a very crucial part. Wang $et \ al.$ proposed LEAF algorithm to solve the trade-off problem between accuracy, energy consumption and latency \cite{wang2020user}. However, this work only consider the limited scenario of one server. Although Ahn $et \ al.$ proposed another algorithm to solve the abovementioned problem \cite{ahn2020novel}, it also only consider the one server scenario. Huang $et \ al.$ considered delay and user location as optimization objectives, but did not consider energy consumption \cite{huang2021proactive}. He $et \ al.$ formulated an excellent model, but only accuracy is inside \cite{he2020optimizing}.\par
Furthermore, none of the works mentioned above consider the energy consumption of edge servers. Unlike cloud servers, edge servers face energy consumption issues as well.

\section{Problem Formulation}\label{Problem Formulation}
In this chapter, we will overview the edge-based MAR system of this paper in Fig. \ref{fig-1}. And the next section describes analytical model which is used to formulate the problem.
\subsection{System Overview}
\begin{figure*}[htbp]
    \centerline{\includegraphics[width=12.0cm]{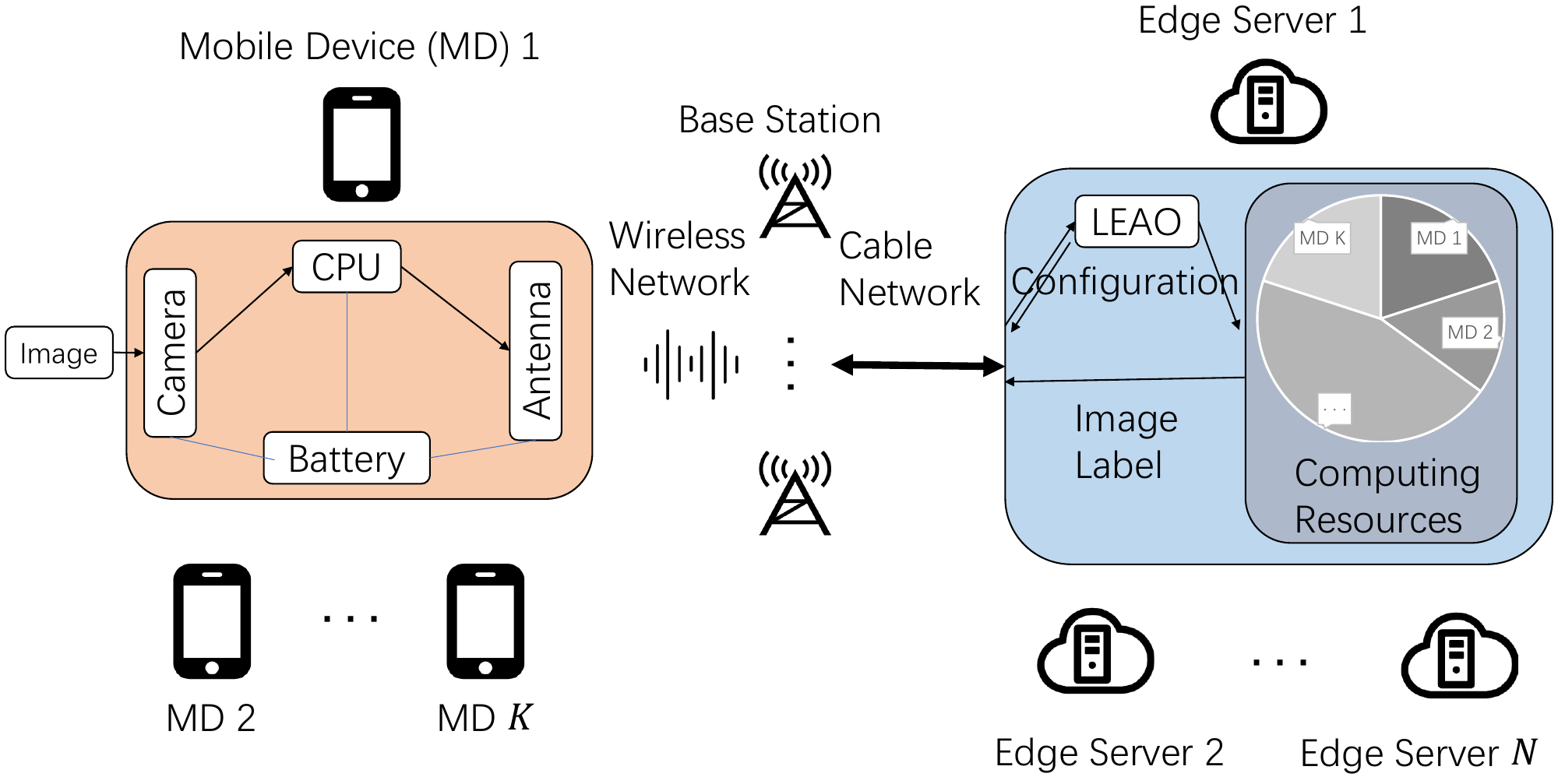}}
\caption{System Overview.}
\vspace{-15pt}
\label{fig-1}
\end{figure*}
In Fig. \ref{fig-1}, there are $K$ mobile devices or users and $N$ edge servers. Each mobile device will choose one and only one edge server to connect and off-load their computing tasks at one time. 
Throughout this paper, the $n$-th dimension of an $N$-dimensional vector $x$ is denoted by $x_n$. We use an indicator $a_{k,n}$ to indicate the connection between mobile devices and servers. If $a_{k,n}=1$, $k$-th mobile device connects with $n$-th server. We use matrix $\boldsymbol{A}$ to denote the set of variables $a_{k,n}$.\par
\vspace{-5pt}
\begin{equation}
\begin{aligned}
    \boldsymbol{A}=[a_{k,n}|_{k \in \mathcal{K}, n\in \mathcal{N}}], \sum_{n \in \mathcal{N}}a_{k,n} = 1\\
    \mathcal{K}:=\{1, ..., K\}, \mathcal{N}:=\{1, ..., N\}
\end{aligned}
\end{equation}

At the same time, the corresponding server will dynamically allocate the computing resources $r_k$ to the $k$-th mobile device, $\boldsymbol{r}:=\{r_1, ..., r_K\}$. The LEAO algorithm executes on the server, monitors the network information, and timely delivers the configuration to both mobile devices and servers.\par
In this paper, we analyze the entire image processing workflow: image generation, data transmission, image processing on servers, and accuracy evaluation. We take into account the system latency, as well as the energy consumption of both mobile devices and servers, and image recognition accuracy. We will formulate the above optimization objective in the following subsection.\par
\subsection{Latency}
Our latency model is constructed according to Equation \eqref{L}, where $L_{k}^{t}$ is the image transmission latency, $L_{k}^{cn}$ is the core network latency and $L_{k}^{p}$ is the processing latency.
\begin{equation}
L_{k}(s_k, a_{k,n},r_k)=L_{k}^{t}(s_k)+L_{k}^{cn}(a_{k,n})+L_{k}^{p}(s_k, r_k)
\label{L}
\end{equation}

We denote the video frame resolution of the $k$-th mobile device as $s_k$, whose unit is pixel and $\sigma$ is the number of bits in one pixel. Then, the transmission latency is modeled as Equation \eqref{L_t} and we have $\boldsymbol{s}:=\{s_1, ..., s_K\}$.

\begin{equation}
L_{k}^{t}(s_k)=\sigma s_k / R_k
\label{L_t}
\end{equation}
where $R_k$ is the wireless data rate of the $k$-th mobile device.

And in Equation \eqref{L_cn}, $l_{k,n}$ is the core network latency between $k$-th mobile device and $n$-th server.
\begin{equation}
L_{k}^{cn}(a_{k,n})= \sum_{n \in \mathcal{N}}a_{k,n} l_{k,n}
\label{L_cn}
\end{equation}
We denote the complexity of task of $k$-th mobile device by $C_k$. Then, the processing latency can be described by Equation \eqref{L_p}, where we assume $C(s_k)$ is a convex function with respect to $s_k$.  

\vspace{-10pt}
\begin{equation}
L_{k}^{p}(s_k, r_k)= C(s_k) / r_k
\label{L_p}
\vspace{-5pt}
\end{equation}

\subsection{Accuracy}
In our system, we regard the image recognition accuracy as one of our optimization goal because it is directly related to user experience. We assume that the accuracy $A_k(s_k)$ is a concave function of video frame resolution $s_k$.


\subsection{Energy Consumption}
Our mobile device energy consumption model is shown in Equation \eqref{for-6}, where $E_{k}^{img}$ is the image generation and preview energy consumption, $E_{k}^{com}$ is the wireless communication energy consumption, $E_{k}^{bs}$ is the base energy consumption of mobile device. We use the sum of them to denote the total energy consumption of the $k$-th mobile device:
\begin{equation}
E_{k}(f_k, s_k, a_{k,n},r_k) = E_{k}^{img} + E_{k}^{com} + E_{k}^{bs}
\label{for-6}
\end{equation}
We denote the CPU frequency of $k$-th mobile device by $f_k$. In general, the most significant proportion of energy consumption is often $E_{k}^{img}$, which is the the product of delay and power:
\begin{equation}
E_{k}^{img}(f_k)= t_{pre} P^{pre}(f_k)
\label{for-7}
\end{equation}
where $t_{pre}$ is the time to pre-process an image and is assumed to be a constant. $P^{pre}(f_k)$ is the power of pre-processing an image and we suppose that it is convex with respect to $f_k$. Finally, we set $E_{k}^{com}$ and $E_{k}^{bs}$ as following equation.
\begin{equation}
E_{k}^{com}(s_k)= P^{tr} (R_k) L_k^t(s_k)
\label{for-8}
\end{equation}
\begin{equation}
E_{k}^{bs}(f_k, s_k, a_{k,n},r_k)=P^{bs} (f_k) L_k(s_k, a_{k,n},r_k)
\label{for-9}
\end{equation}

where $P^{tr} (R_k)$ is the transmission power of $k$-th mobile device and $P^{bs} (f_k)$ is the basic power of mobile device. We assume that $P^{bs} (f_k)$ is a convex function with respect to $f_k$. 
Furthermore, the server energy consumption is modeled as Equation \eqref{for-13}.
\begin{equation}
E_n(a_{k,n}, s_k, r_k) = \sum_{k \in \mathcal{K}} a_{k,n} P_n(\frac{r_k}{S_n}F_n)L_k^p(s_k, r_k)
\label{for-13}
\end{equation}
In this equation, $F_n$ is the CPU frequency of $n$-th server and $P_n$ is the power of $n$-th server. For each server $n$, only the computing resources that are used will generate energy consumption and $S_n$ stands for the total available resources of $n$-th server. Finally, the argument of the $P_n$ function is obtained by multiplying the proportion
of computing resources used by each server $r_k/S_n$ and the CPU frequency $F_n$.

\subsection{Optimization problem}
In this paper, we aim to minimize the overall energy consumption, latency and maximize the accuracy of each mobile device. This is a multi-objective optimization problem and we adopt the weighted sum method to express the objective function $Q$ given in the Equation \eqref{obj_func}. Furthermore, to express the trade-off between different objectives, we introduce parameter $\lambda_{1}^{k}$ and $\lambda_{2}^{k}$ to reflect the preference between them. For example, a larger $\lambda_{1}^{k}$ indicates that the system prefers a lower latency and a larger $\lambda_{2}^{k}$ indicates that the system prefers the higher accuracy.\par
\vspace{-15pt}
\begin{equation}\label{obj_func}
    Q(f_k, s_k, a_{k,n},r_k) \!= \! \sum_{n\in \mathcal{N}}\textstyle{\frac{E_n}{N}}
    \!+\! \sum_{k\in \mathcal{K}} \textstyle{\frac{E_{k} + \lambda_{1}^{k}L_{k}-\lambda_{2}^{k}A_{k}}{K}}
\end{equation}

Besides, the optimization problem is shown in equation \eqref{problem_p0}.\par
\vspace{-15pt}

\begin{subequations}\label{problem_p0}
\begin{align}
\mathbb{P}_{0}: & \min_{\{\boldsymbol{f}, \boldsymbol{s}, \boldsymbol{A}, \boldsymbol{r} \}} Q \tag{\ref{problem_p0}} \\
 s.t. \quad &C_1:  A_k(s_k) \ge A_{min}, \forall k\in \mathcal{K}; \label{cons_p0_A}\\
& \textstyle{C_2:  L^{k}(s_k, a_{k,n},r_k)\leq L_{max}^k, \forall k\in \mathcal{K}; }\label{cons_p0_L}\\
& \textstyle{C_3:  f_{min}\leq f_k \leq f_{max}, \forall k\in \mathcal{K};}\label{cons_p0_f} \\
& \textstyle{C_4:  s_{min}\leq s_k \leq s_{max}, \forall k\in \mathcal{K};} \label{cons_p0_s}\\
& \textstyle{C_5:  a_{k,n} \in {0,1}, \forall k\in \mathcal{K}, n \in \mathcal{N};}\label{cons_p0_a1}\\
& \textstyle{C_6: \sum_{n \in \mathcal{N}}a_{k,n} = 1, \forall k\in \mathcal{K}; }\label{cons_p0_a2}\\
& \textstyle{C_7:  \sum_{k\in \mathcal{K}}a_{k,n}r_k \leq S_n, \forall n\in \mathcal{N};} \label{cons_p0_ar}
\end{align}
\end{subequations}

where $A_{min}$ is the minimum accuracy requirement of mobile device; $L_{max}^k$ is the upper bound for the $k$-th mobile device latency. Constraints \eqref{cons_p0_f} and \eqref{cons_p0_s} are the constraints of the mobile device’s CPU frequency and pixels of input images. \eqref{cons_p0_a1} and \eqref{cons_p0_a2} ensure that an mobile device only can choose one edge server. \eqref{cons_p0_ar} ensure that the total computing resources allocated to mobile devices connected to $n$-th server do not exceed the computing resources of $n$-th server $S_n$.\par
$\mathbb{P}_{0}$ is MINP which is a NP-hard problem. To solve such an intractable problem, we analyze the properties of $\mathbb{P}_{0}$ firstly. In the objective function $Q$, the four variables $\boldsymbol{f}$, $\boldsymbol{s}$, $\boldsymbol{A}$ and $\boldsymbol{r}$ are all multiplying to each other. Therefore, it is obvious that $\mathbb{P}_{0}$ is not jointly convex with respect to $[ \boldsymbol{f}, \boldsymbol{s}, \boldsymbol{A}, \boldsymbol{r} ]$. Similarly, it is easy to prove that $\mathbb{P}_{0}$ is strictly convex with respect to $\boldsymbol{f}$, $\boldsymbol{s}$, and $\boldsymbol{r}$ separately since the second-order derivative of $Q$ with respect to each of them are greater than zero. Due to the simplicity of this proof and the limitation of the length, a detailed description will not be provided here. Although $\mathbb{P}_{0}$ is not jointly convex with respect to $[ \boldsymbol{f}, \boldsymbol{s}, \boldsymbol{A}, \boldsymbol{r} ]$, we can use other problems to approximate the solution of $\mathbb{P}_{0}$, which will be explained in Section \ref{Algorithm}.

\section{Optimization Algorithm}\label{Algorithm}
The first difficulty in solving the problem $\mathbb{P}_{0}$ arises from the discrete variable $\boldsymbol{A}$. We relax the discrete variable $a_{k,n}$ into
continuous variable $\hat{a}_{k,n}$, $\boldsymbol{\hat{A}}=[\hat{a}_{k,n}|_{k \in \mathcal{K}, n\in \mathcal{N}}]$. By changing constraints \eqref{cons_p0_a1} and \eqref{cons_p0_a2} of $\mathbb{P}_{0}$ into \eqref{cons_p1_a1}, \eqref{cons_p1_a2} and \eqref{cons_p1_a3} of $\mathbb{P}_{1}$, we transform $\mathbb{P}_{0}$ into an equivalent $\mathbb{P}_{1}$ since variable $\hat{a}_{k,n}$ also only can be 0 or 1.\par

\vspace{-10pt}
\begin{subequations}\label{problem_p1}
    \begin{align}
    \mathbb{P}_{1}: & \min_{\{\boldsymbol{f}, \boldsymbol{s}, \boldsymbol{\hat{A}}, \boldsymbol{r} \}} Q \tag{\ref{problem_p1}} \\
     s.t. \quad &\eqref{cons_p0_A}, \eqref{cons_p0_L}, \eqref{cons_p0_f}, \eqref{cons_p0_s}; \notag\\
    & \textstyle{C_5:  0\leq \hat{a}_{k,n} \leq 1, \forall n\in \mathcal{N}, \forall k\in \mathcal{K};} \label{cons_p1_a1}\\
    & \textstyle{C_6:  \sum_{k\in \mathcal{K}} \sum_{n\in \mathcal{N}} \hat{a}_{k,n}(1-\hat{a}_{k,n}) \leq 0;} \label{cons_p1_a2}\\
    & \textstyle{C_7: \sum_{n \in \mathcal{N}}\hat{a}_{k,n} = 1, \forall k\in \mathcal{K};} \label{cons_p1_a3}\\
    & \textstyle{C_8:  \sum_{k\in \mathcal{K}}\hat{a}_{k,n}r_k \leq S_n, \forall n\in \mathcal{N};} \label{cons_p1_ar}
    \end{align}
\end{subequations}
Since we only modify the constraints related to $\boldsymbol{A}$, $\mathbb{P}_{1}$ is also strictly convex with respect to $\boldsymbol{f}$, $\boldsymbol{s}$ and $\boldsymbol{r}$ and all the variables in $\mathbb{P}_{1}$ are continuous. However, it is still difficult to solve for the nonconvex part. The constraint \eqref{cons_p1_a2} in $\mathbb{P}_{1}$ is concave so that $\mathbb{P}_{1}$ is not convex with respect to $\boldsymbol{\hat{A}}$, and $\mathbb{P}_{1}$ is also not jointly convex with respect to $[ \boldsymbol{f}, \boldsymbol{s}, \boldsymbol{\hat{A}}, \boldsymbol{r} ]$. Within this section, a systematic algorithm will be formulated step by step to address and resolve the problem $\mathbb{P}_{1}$.

\subsection{Successive Convex Approximation (SCA) Algorithm}
We plan to use SCA to solve the nonconvex part. However, to facilitate the solution, we need to penalize the concave constraint in \eqref{cons_p1_a2} to the objective function before we use SCA, which is shown in \eqref{problem_p2}.\par

\vspace{-10pt}
\begin{subequations}\label{problem_p2}
    \begin{align}
    \mathbb{P}_{2}: & \min_{\{\boldsymbol{f}, \boldsymbol{s}, \boldsymbol{\hat{A}}, \boldsymbol{r} \}} Q - \mu \sum_{k\in \mathcal{K}} \sum_{n\in \mathcal{N}} \hat{a}_{k,n}(\hat{a}_{k,n} - 1) \tag{\ref{problem_p2}} \\
     s.t. \quad &\eqref{cons_p0_A}, \eqref{cons_p0_L}, \eqref{cons_p0_f}, \eqref{cons_p0_s}, \eqref{cons_p1_a1}, \eqref{cons_p1_a3}, \eqref{cons_p1_ar}; \notag
    \end{align}
\end{subequations}

$\mu$ in $\mathbb{P}_{2}$ is the penalty parameter and we have $\mu \ge 0$. Denote with $\alpha(\mu)$ the optimal objective value. Based on Theorem 1 of \cite{le2012exact}, we show the equivalence of $\mathbb{P}_{1}$ and $\mathbb{P}_{2}$ in the following lemma.\par

\vspace{-8pt}
\begin{lemma}
    (Exact Penalty) For all $\mu \ge \mu_0$ where 
    \begin{equation}
        \mu_0 = \frac{Q (f_k, s_k, a_{k,n}^0,r_k) - \alpha(0)}{max_{\boldsymbol{\hat{A}}} \{ \hat{a}_{k,n}(\hat{a}_{k,n} - 1): \eqref{cons_p1_a1}, \eqref{cons_p1_a2}, \eqref{cons_p1_a3} \}}
    \end{equation}
    With any $a_{k,n}^0, \forall n\in \mathcal{N}, \forall k\in \mathcal{K}$ satisfying constraints \eqref{cons_p1_a1}, \eqref{cons_p1_a2} and \eqref{cons_p1_a3}, $\mathbb{P}_{1}$ and $\mathbb{P}_{2}$ have the same optimal solution.
    \begin{proof}
        Theorem 1 in \cite{le2012exact} has proved this lemma. And the values of $f_k$, $s_k$ and $r_k$ are from last iteration.
    \end{proof}
\label{le1}
\end{lemma}
SCA involves iteratively solving a sequence of convex problems. In each iteration, we use a surrogate convex function to approximate the non-convex function. Specifically, we approximate $\mathbb{P}_{2}$ with $\mathbb{O}_{t}$ which is the problem in $t$-th iteration shown in following equation. In $\mathbb{O}_{t}$, we approximate constraints \eqref{cons_p1_a2} with $\hat{a}_{k,n}^{(t)}(\hat{a}_{k,n}^{(t)} - 1) + (2\hat{a}_{k,n}^{(t)} - 1)(\hat{a}_{k,n} - \hat{a}_{k,n}^{(t)})$, and $t=:\{0,1,...\}$. At the end of one iteration, we get the value of $\hat{a}_{k,n}$ and we regard it as the $\hat{a}_{k,n}^{(t+1)}$ in next iteration. This part is as shown in Algorithm \ref{alg_SCA}.\par

\vspace{-10pt}

\begin{subequations}\label{prblem_ot}
    \begin{align}
    \mathbb{O}_{t}: & \min_{\{\boldsymbol{f}, \boldsymbol{s}, \boldsymbol{\hat{A}}, \boldsymbol{r} \}} Q - \mu \sum_{k\in \mathcal{K}} \sum_{n\in \mathcal{N}} (\hat{a}_{k,n}^{(t)}(\hat{a}_{k,n}^{(t)} - 1) + \notag \\
    & \quad \quad \quad (2\hat{a}_{k,n}^{(t)} - 1)(\hat{a}_{k,n} - \hat{a}_{k,n}^{(t)})) \tag{\ref{prblem_ot}} \\
    s.t. \quad &\eqref{cons_p0_A}, \eqref{cons_p0_L}, \eqref{cons_p0_f}, \eqref{cons_p0_s}, \eqref{cons_p1_a1}, \eqref{cons_p1_a3}, \eqref{cons_p1_ar}; \notag
    \end{align}
\end{subequations}

\vspace{-20pt}
\begin{figure}[htbp] 
        \renewcommand{\algorithmicrequire}{\textbf{Input:}}
        \renewcommand{\algorithmicensure}{\textbf{Output:}}
        \begin{algorithm}[H]
            \caption{\label{alg_SCA} SCA}
            \begin{algorithmic}[1]
                \REQUIRE optimization variable $\boldsymbol{x}$, objective function $O(\boldsymbol{x} | \boldsymbol{x}^{(t)})$, constraints set $\boldsymbol{C}$ and convergence condition $\tau$ \par
                \ENSURE optimal solution $\boldsymbol{x}^*$\par
                \STATE Initialization: Find an initial feasible point $\boldsymbol{x}^{(0)}$ of $\mathbb{P}_{2}$ and set t = 0.

                \FOR{iteration $t$ = $0,1,..$}
                    \STATE $\boldsymbol{x}^{(t+1)} \gets$ solve $O(\boldsymbol{x} | \boldsymbol{x}^{(t)})$ satisfying $\boldsymbol{C}$
                    \IF{$|(\boldsymbol{x}^{(t+1)} - \boldsymbol{x}^{(t)})/\boldsymbol{x}^{(t)}| \leq \tau$}
                        \STATE \textbf{break;}
                    \ENDIF
                    \STATE $t \gets t + 1$  
                \ENDFOR
                \STATE $\boldsymbol{x}^* \gets \boldsymbol{x}^{(t)}$
            \end{algorithmic}
        \end{algorithm}
\vspace{-0.4cm}
\end{figure}
From $\mathbb{P}_{2}$ to $\mathbb{O}_{t}$, we didn't change any constraints or other part of objective function. Therefore, the problem $\mathbb{O}_{t}$ is strictly convex with respect to $\boldsymbol{f}$, $\boldsymbol{s}$, and $\boldsymbol{r}$ separately. Furthermore, it's easy to prove that the problem $\mathbb{O}_{t}$ is also strictly convex with respect to $\boldsymbol{\hat{A}}$ due to the linearity of problem $\mathbb{O}_{t}$ with respect to $\boldsymbol{\hat{A}}$. Due to the limitation of the length, a detailed description will not be provided here.\par



However, at this stage, the problem is still not completely solved, as the problem $\mathbb{O}_{t}$ is not jointly convex in all variables, because there are many products of variables $\boldsymbol{f}, \boldsymbol{s}, \boldsymbol{\hat{A}}$, and $\boldsymbol{r}$ in the objective function and the constraints. We next introduce new variables and use Product Replacement to further address this problem.\par

\subsection{Product Replacement Algorithm}

To address the non-joint convexity of the problem $\mathbb{O}_{t}$ with respect to $[ \boldsymbol{f}, \boldsymbol{s}, \boldsymbol{\hat{A}}, \boldsymbol{r} ]$, we introduce new variables $\boldsymbol{w}:=\{w_1, ..., w_K\}$ and $\boldsymbol{z}:=\{z_1, ..., z_N\}$ to separate the product of $\boldsymbol{s}$ and $\boldsymbol{r}$, as well as $\boldsymbol{\hat{A}}$ and $\boldsymbol{r}$ according to the algorithm in \cite{zhao2023human}.\par

The Section \uppercase\expandafter{\romannumeral4} in \cite{zhao2023human} has proved that the objective function $xy$ has the same KKT solution as $x^2w+y^2/4w$ when $w=y/2x$. Then we apply this method in formula \eqref{L_p} and we get new processing delay $\widetilde{L}_{k}^{p}(s_k, r_k, w_k)$. The new objective function is denoted by $\widetilde{Q}$.\par

\vspace{-5pt}
\begin{equation}
    \widetilde{L}_{k}^{p}(s_k, r_k, w_k)= C^2(s_k)w_k + \frac{1}{4 w_k r^2_k}
\end{equation}

At the same time, this method also can be applied in constraint \eqref{cons_p1_ar}, then we get the new constraint \eqref{cons_ht_ar} in following problem.\par

\vspace{-15pt}

\begin{subequations}\label{prblem_ht}
    \begin{align}
    \mathbb{H}_{t}: & \min_{\{\boldsymbol{f}, \boldsymbol{s}, \boldsymbol{\hat{A}}, \boldsymbol{r}, \boldsymbol{w}, \boldsymbol{z} \}} \widetilde{Q} \tag{\ref{prblem_ht}} \\
    s.t. \quad &\eqref{cons_p0_A}, \eqref{cons_p0_L}, \eqref{cons_p0_f}, \eqref{cons_p0_s}, \eqref{cons_p1_a1}, \eqref{cons_p1_a3}; \notag \\
    &C_7: \sum_{k\in \mathcal{K}}(\hat{a}_{k,n}^2z_n + \frac{r^2_k}{4z_n}) \leq S_n, \forall n\in \mathcal{N}; \label{cons_ht_ar}
    \end{align}
\end{subequations}
Via the proof in the Section \uppercase\expandafter{\romannumeral4} in \cite{zhao2023human}, we can draw the conclusion that the problem $\mathbb{H}_{t}$ and the problem $\mathbb{O}_{t}$ have the same KKT solution for variables $[\boldsymbol{f}, \boldsymbol{s}, \boldsymbol{\hat{A}}, \boldsymbol{r}]$.\par
However, the product of $f_k$ and $\hat{a}_{k,n}$, $s_k$ and $r_k$ still exists in the objective function of $\mathbb{H}_{t}$. Introducing new variables to split these products would lead to a significant expansion of the variable space. To avoid the problem becoming overly complex, we treat $\boldsymbol{f}$ as a constant and optimize only $[\boldsymbol{s}, \boldsymbol{\hat{A}}, \boldsymbol{r}, \boldsymbol{w}, \boldsymbol{z}]$. This process is presented in Algorithm \ref{alg_AO} where some operation symbols are from MATLAB.\par

\begin{figure}[htbp] 
        \renewcommand{\algorithmicrequire}{\textbf{Input:}}
        \renewcommand{\algorithmicensure}{\textbf{Output:}}
        \begin{algorithm}[H]
            \caption{\label{alg_AO} Product Replacement}
            \begin{algorithmic}[1]
                \REQUIRE Problem $\mathbb{H}_{t}$ and convergence condition $\tau$ \par
                \ENSURE optimal solution $[\boldsymbol{s}, \boldsymbol{\hat{A}}, \boldsymbol{r}, \boldsymbol{w}, \boldsymbol{z}]^*$\par
                \STATE Initialization: Find an initial point of $[\boldsymbol{s}, \boldsymbol{\hat{A}}, \boldsymbol{r}, \boldsymbol{w}, \boldsymbol{z}]$.

                \FOR{iteration $i$ = $0,1,..$}
                    \STATE $[\boldsymbol{s}, \boldsymbol{\hat{A}}, \boldsymbol{r}]^{(i)} \gets$ use Algorithm \ref{alg_SCA} to solve $\mathbb{H}_{t}$ with fixed $[\boldsymbol{f}, \boldsymbol{w}, \boldsymbol{z}]$
                    \STATE $\boldsymbol{w}^{(i)} = 1 ./ (2C(\boldsymbol{s}^{(i)}) .* \boldsymbol{r}^{(i)})$
                    \STATE $\boldsymbol{z}^{(i)} = sum(\boldsymbol{r}^{(i)}) ./ 2sum(\boldsymbol{\hat{A}}^{(i)}, 1)$
                    \IF{$|(\widetilde{Q}^{(i+1)} - \widetilde{Q}^{(i)})/\widetilde{Q}^{(i)}| \leq \tau$}
                        \STATE \textbf{break;}
                    \ENDIF
                    \STATE $i \gets i + 1$  
                \ENDFOR
                \STATE $[\boldsymbol{s}, \boldsymbol{\hat{A}}, \boldsymbol{r}, \boldsymbol{w}, \boldsymbol{z}]^* \gets [\boldsymbol{s}, \boldsymbol{\hat{A}}, \boldsymbol{r}, \boldsymbol{w}, \boldsymbol{z}]^{(i)}$
            \end{algorithmic}
        \end{algorithm}
\vspace{-25pt}
\end{figure}

\subsection{LEAO}
In this subsection, we use Block Coordinate Descent (BCD) to optimize $\boldsymbol{f}$ which was kept constant in last subsection. Firstly, we fix the $\boldsymbol{f}$ and optimize $[\boldsymbol{s}, \boldsymbol{\hat{A}}, \boldsymbol{r}, \boldsymbol{w}, \boldsymbol{z}]$ with Algorithm \ref{alg_SCA} and Algorithm \ref{alg_AO}. Secondly, we fix $[\boldsymbol{s}, \boldsymbol{\hat{A}}, \boldsymbol{r}, \boldsymbol{w}, \boldsymbol{z}]$ and optimize $\boldsymbol{f}$ using the KKT condition.\par
Combining all the above lemma and algorithms, we are finally able to design the algorithm LEAO, which is shown in the Algorithm \ref{alg:LEAO}. In the first step, we employ BCD to optimize $\boldsymbol{f}$ separately from the other variables. Then, we use SCA to ensure that the optimization problem is convex with respect to $\boldsymbol{\hat{A}}$. Finally, we introduce new variables to separate the product of $\boldsymbol{s}$ and $\boldsymbol{r}$, as well as $\boldsymbol{\hat{A}}$ and $\boldsymbol{r}$, such that the problem becomes jointly convex with respect to $[\boldsymbol{s}, \boldsymbol{\hat{A}}, \boldsymbol{r}]$.

\vspace{-15pt}
\begin{figure}[htbp] 
        \renewcommand{\algorithmicrequire}{\textbf{Input:}}
        \renewcommand{\algorithmicensure}{\textbf{Output:}}
        \begin{algorithm}[H]
            \caption{\label{alg:LEAO} LEAO}
            \begin{algorithmic}[1]
                \REQUIRE problem $\mathbb{P}_{1}$, problem $\mathbb{H}_{t}$, convergence condition $\tau$\par
                \ENSURE $[\boldsymbol{f}, \boldsymbol{s}, \boldsymbol{\hat{A}}, \boldsymbol{r}]^*$\par
                \STATE $\boldsymbol{\hat{A}} \gets$ $\boldsymbol{A}$

                \FOR{iteration $i$ = $1,2,..$}
                    \STATE $\boldsymbol{f}^{(i)} \gets$  use KKT condition to solve $\mathbb{P}_{1}$ with fixed $[\boldsymbol{s}, \boldsymbol{\hat{A}}, \boldsymbol{r}]$
                    \STATE $[\boldsymbol{s}, \boldsymbol{\hat{A}}, \boldsymbol{r}]^{(i)} \gets$ use Algorithm \ref{alg_AO} to solve $\mathbb{H}_{t}$ with fixed $\boldsymbol{f}$
                    \IF{$|(Q^{(i+1)} - Q^{(i)})/Q^{(i)}| \leq \tau$}
                        \STATE \textbf{break;}
                    \ENDIF
                    \STATE $i \gets i + 1$  
                \ENDFOR
                \STATE $[\boldsymbol{f}, \boldsymbol{s}, \boldsymbol{\hat{A}}, \boldsymbol{r}]^* \gets [\boldsymbol{f}, \boldsymbol{s}, \boldsymbol{\hat{A}}, \boldsymbol{r}]^{(i)}$
            \end{algorithmic}
        \end{algorithm}
\vspace{-20pt}
\end{figure}

\subsection{Convergence and Time Complexity Analysis}
The convergence proof of SCA, Product Replacement and BCD can be found in \cite{razaviyayn2014successive, zhao2023human, beck2013convergence}. In view of the fact that the main body of Algorithm LEAO is composed of these three algorithms, its convergence is guaranteed as well.\par
Since the objective function $\widetilde{Q}$ is jointly convex with respect to $[\boldsymbol{s}, \boldsymbol{\hat{A}}, \boldsymbol{r}]$, we can solve it by KKT condition. Therefore, we assume the time complexity of this process is $O(1)$. Since the size of the variable space is $3K+N+KN$, the time complexity of solving this problem is $O(KN)$. Assuming the loop counts of the three algorithms mentioned above are $\zeta$, $\eta$, $\theta$, the time complexity of the final algorithm LEAO is $O(\zeta \eta \theta KN)$.

\section{Performance Evaluation}\label{simulation}
In our simulation, the default configuration is 100 users and 10 servers and we generate $l_{k,n}$ randomly. Different parameters default setting are showed in Table \ref{tab:opt}. Besides, the parameter $\mu$ in Lemma \ref{le1} is set by experience.\par

\vspace{-5pt}
\begin{table}[htbp]
    \centering
    \caption{Parameters Setting}
    \begin{tabular}{|l|c|l|c|}
    \hline
    Parameters & Value & Parameters & Value \\
    \hline
    $\tau$ & 0.001 & $\lambda_1^k$ & 0 $\sim$ 50 \\
    \hline
    $\sigma$ & 8 bits & $\lambda_2^k$ & 0 $\sim$ 1000 \\
    \hline
    $S_n$ & 8 $\sim$ 12 TFLOPS & $f_{min}$ & 2.2 GHz\\
    \hline
    $F_n$ & 4.4GHz $\sim$ 4.6GHz & $f_{max}$ & 3.5 GHz\\
    \hline
    $t_{pre}$ & 0.4 ms & $s_{min}$ & $256^2$ pixels\\
    \hline
    $A_{min}$ & 0.6 & $s_{max}$ & $1024^2$ pixels\\
    \hline
    $l_{k,n}$ & 100ms $\sim$ 130ms & $L_{max}$ & 250 ms\\
    \hline
    \end{tabular}
    \label{tab:opt}
\end{table}

Based on the measurements of other works \cite{liu2018edge, wang2020user}, a list of specific function used in this paper is shown in Table \ref{tab:model}. Although the function $P^{pre}(f_k)$ is not a convex function, it's convex when $f_k$ is in the range of Table \ref{tab:opt}.\par

\vspace{-5pt}
\begin{table}[htbp]
    \centering
    \caption{proposed models}
    \begin{tabular}{|l|c|}
    \hline
    Functions & Models \\
    \hline
    $C(s_k)$ & $7 \times 10^{-10}s_k^{1.5}+0.083$ TFLOPS\\
    \hline
    $A(s_k)$ & $1-1.578e^{-6.5 \times 10^{-3} \sqrt{s_k}}$\\
    \hline
    $P^{tr}(R_k)$ & $0.018R_k + 0.7$ \\
    \hline
    $P^{bs}(f_k)$ & $0.079f_k + 0.59$ \\
    \hline
    $P^{pre}(f_k)$ & $-0.01071f_k^3 + 0.06055f_k^2 - 0.1028f_k + 0.107$ \\
    \hline
    $P_n(x)$ & $0.083x^2 + 0.32$\\
    \hline
    \end{tabular}
    \label{tab:model}
\end{table}

Furthermore, in this paper, we compare LEAO with three algorithms.
\begin{itemize}
    \item \textbf{Baseline}: The baseline algorithm has fixed variable values and we generate those values randomly by the same Gaussian random seed and the variable range is shown in Table \ref{tab:opt}.
    \item \textbf{User Workload Optimized (UWO)}: This algorithm optimize mobile device's CPU frequency $\boldsymbol{f}$ and image resolution $\boldsymbol{s}$. All other variable are fixed.
    \item \textbf{Resources Allocation Optimized (RAO)}: This algorithm optimize the resources allocation of mobile devices $\boldsymbol{r}$ and the connection map $\boldsymbol{A}$. And other variable are fixed.
\end{itemize}

\begin{figure}[htbp]
    \begin{minipage}{0.48\linewidth}
        \centerline{\includegraphics[width=4.0cm]{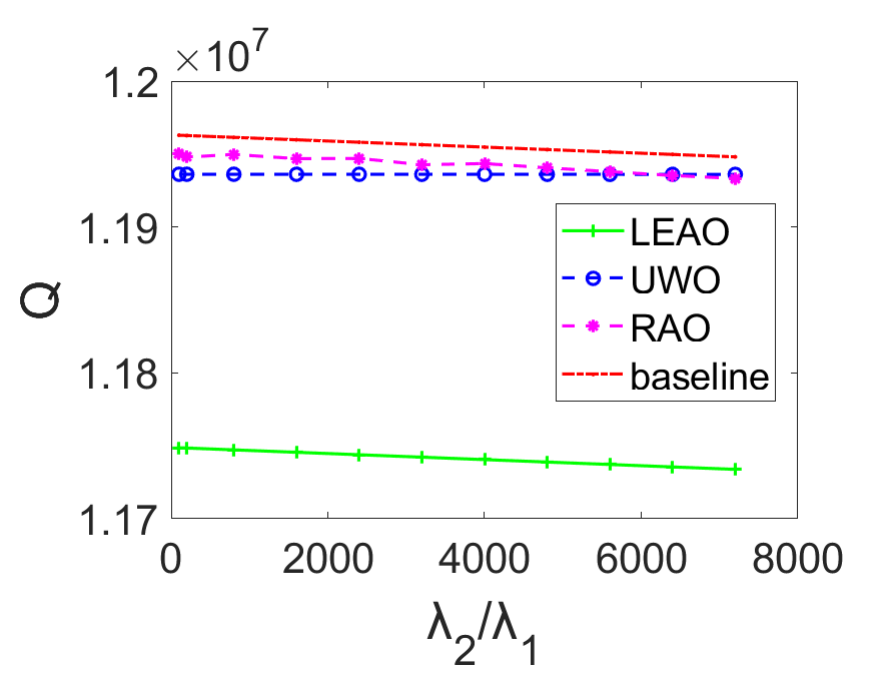}}
        \centerline{(a) Optimality vs. $\lambda_2 / \lambda_1$}
    \end{minipage}
    \hfill
    \begin{minipage}{0.48\linewidth}
        \centerline{\includegraphics[width=4.0cm]{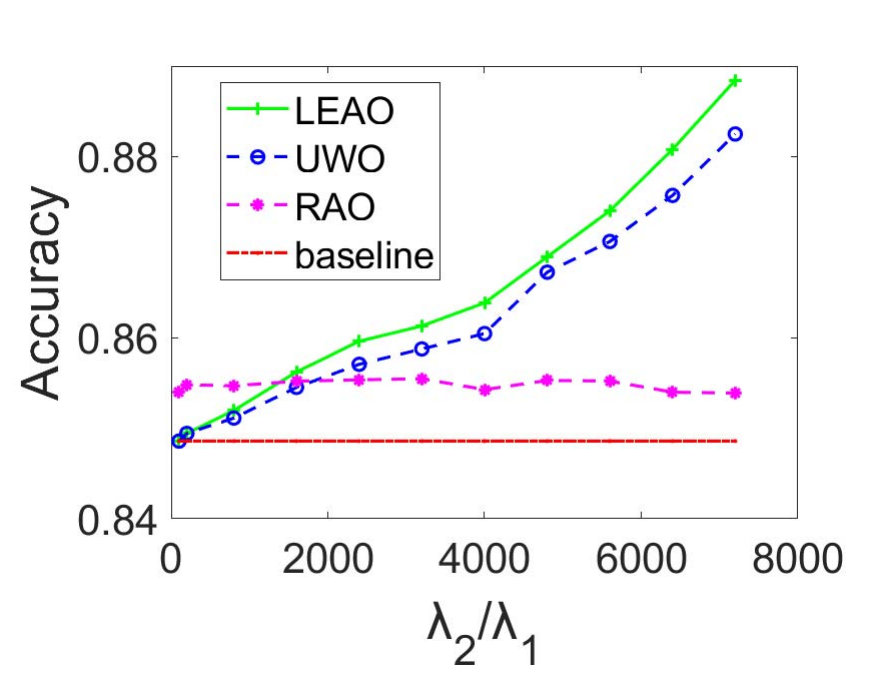}}
        \centerline{(b) Accuracy vs. $\lambda_2 / \lambda_1$}
    \end{minipage}
    \vfill
    \begin{minipage}{0.48\linewidth}
        \centerline{\includegraphics[width=4.0cm]{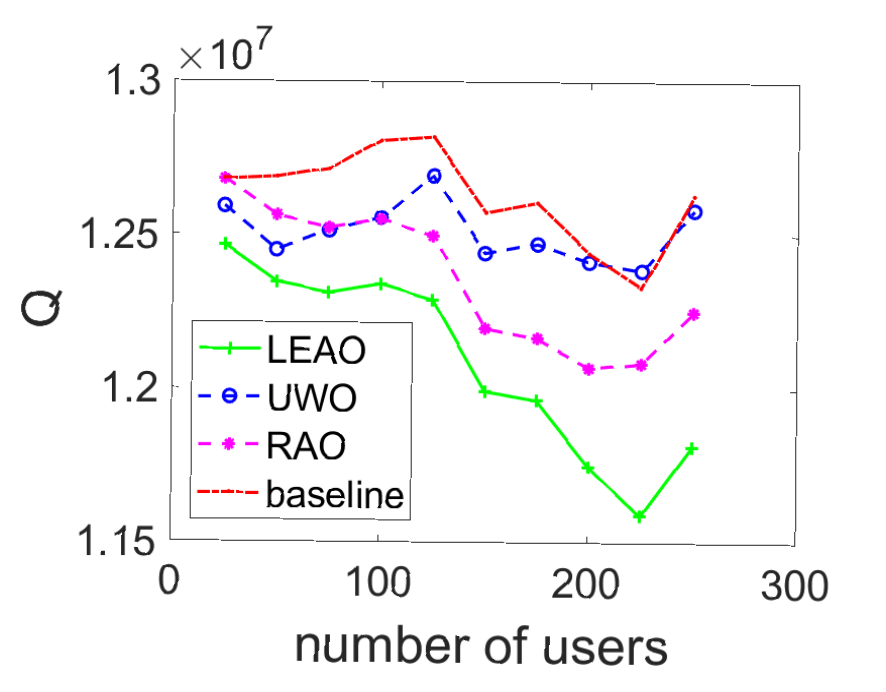}}
        \centerline{(c) Optimality vs. $K$}
    \end{minipage}
    \hfill
    \begin{minipage}{0.48\linewidth}
        \centerline{\includegraphics[width=4.0cm]{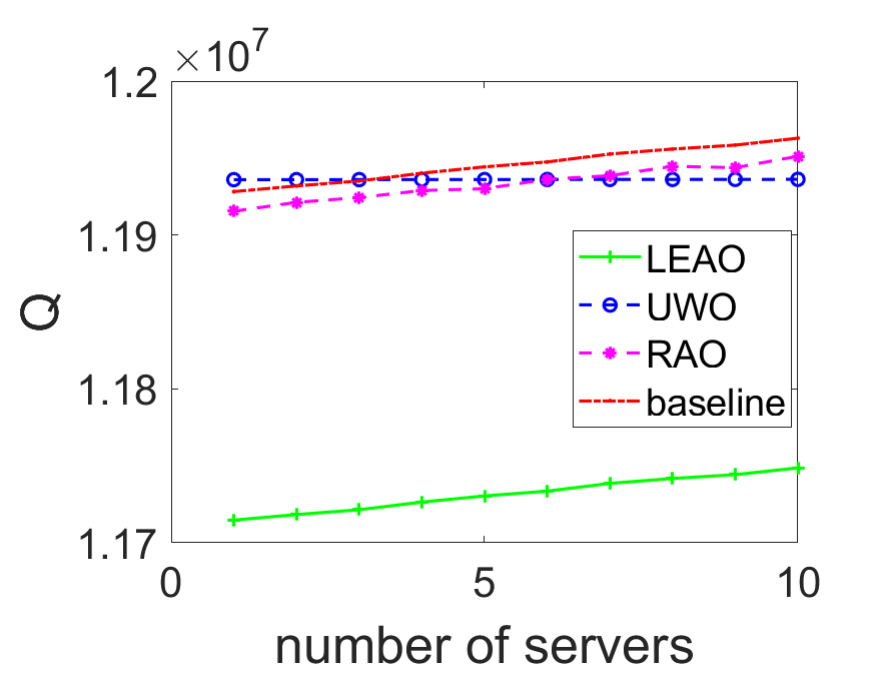}}
        \centerline{(d) Optimality vs. $N$}
    \end{minipage}
    \caption{The influence of different parameters on the experimental results.}
    \label{fig-opt}
\vspace{-10pt}
\end{figure}

\textbf{Optimality.} Firstly, we compare the objective function value between different algorithm and the result is shown in Fig.  \ref{fig-opt}. In one word, LEAO is almost the lowest one, independently of the parameter configuration. Fig. \ref{fig-opt} (a) shows the relationship between $Q$ and $\lambda_2/\lambda_1$. Since accuracy $A_k$ is much smaller number than latency $L_k$, we only increase the $\lambda_2$ and $Q$ decreases with it. Besides, the RAO is better than UWO, which means optimizing the resources allocation is more important than user workload.\par

\textbf{Impact of $\lambda$.} In Fig. \ref{fig-opt} (b), the accuracy increases while the ratio $\lambda_2/\lambda_1$ increasing. Because the larger $\lambda_2$ will make the system pay more attention to the accuracy optimization. Vice versa, the system will focus more on the latency and energy consumption optimization while the $\lambda_2$ is a small number. Simultaneously, with the alteration of the value of the ratio $\lambda_2/\lambda_1$, the objective function associated with the optimal solution exhibits minimal variation, thereby demonstrating the stability of our optimization algorithm's performance.\par

\textbf{Impact of $K$ and $N$.} In Fig. \ref{fig-opt} (c), the objection function of LEAO keeps stable and even decreases as $K$ increases indicating that our algorithm can adapt well to high user volume scenarios. Moreover, the fact that LEAO outperforms RAO implies that, as the number of users increases, optimizing only the resource allocation is not sufficient, and optimizing the users workload is equally important. And in Fig. \ref{fig-opt} (d), the objective function value also slightly decreases as $N$ increases, which indicates that the resource allocation optimization problem with multiple servers is still within the scope of our algorithm's capability.\par

\section{Conclusion}\label{Conclusion}
In this paper, we proposed an edge-based MAR system for Metaverse, which can reduce energy consumption and optimize resources allocation while maintaining high accuracy at the same time. Besides, we built a complete mathematical model to analyze the trade-off between latency, accuracy, energy consumption and resources allocation in edge-based MAR system. On this basis, we developed the LEAO algorithm to improve system performance and user experience by synchronously optimizing mobile device's CPU frequency, frame resolution, server assignment and server resources allocation. Finally, we evaluated the performance of LEAO algorithm by simulation and demonstrated its ability to achieve good experimental results.

\section*{Acknowledgement}
This research is supported by the National Research Foundation, Singapore under its Strategic Capability Research Centres Funding Initiative, the Singapore Ministry of Education Academic Research Fund under Grant Tier 1 RG90/22, Grant Tier 1 RG97/20, Grant Tier 1 RG24/20; and partly by the Nanyang Technological University (NTU)-Wallenberg AI, Autonomous Systems and Software Program (WASP) Joint Project. Any opinions, findings and conclusions or recommendations expressed in this material are those of the author(s) and do not reflect the views of National Research Foundation, Singapore.





\bibliographystyle{IEEEtran}
\bibliography{IEEEbib}

\end{document}